\documentclass[11pt]{article}
\usepackage{vietnam,epsfig}

\def\Journal#1#2#3#4{{#1} {\bf #2}, #3 (#4)}

\def\PRL{\em Phys. Rev. Lett.}

\def\APJ{\em Astrophys. J.}
\def\APJS{\em Astrophys. J. Supp.}
\def\MNRAS{\em Mon. Not. R. Astron. Soc.}

\def\be{\begin{equation}}
\def\ee{\end{equation}}
\def\bea{\begin{eqnarray}}
\def\eea{\end{eqnarray}}

\begin{document}
\vspace*{4cm} \title{PROBING THE ORIGINS OF VOIDS WITH THE CMB}

\author{L. M. Ord$^{\rm A}$, M. Kunz$^{\rm B, \, C}$, H. Mathis$^{\rm D}$ and J. Silk$^{\rm D}$}

\address{$^{\rm A}$\,Department of Astrophysics \& Optics, School of Physics, University of New South Wales, Sydney, NSW 2052, Australia\\
$^{\rm B}$\,Astronomy Centre, University of Sussex, Brighton BN1 9QJ, UK\\
$^{\rm C}$\,Theoretical Physics, University of Geneva, 1211 Geneva 4, Switzerland\\
$^{\rm D}$\,Astrophysics, University of Oxford, Denys Wilkinson Building, Keble Road, Oxford OX1 3RH, UK}

\maketitle\abstracts{ In this talk presented at the 5th Rencontres du
Vietnam 2004, we discuss our preliminary investigations into voids of
primordial origin. We show that if voids in the cold dark matter
distribution existed at the epoch of decoupling, they could contribute
significantly to the apparent rise in cosmic microwave background
(CMB) power on small scales detected by the Cosmic Background Imager
(CBI) Deep Field. Here we present the preliminary results of our
improved method for predicting the effects of primordial voids on the
CMB in which we treat the voids as an external source in the cold dark
matter (CDM) distribution, employing a Boltzmann solver. Our improved
predictions include the effects of a cosmological constant ($\Lambda$)
and acoustic oscillations generated by voids at early times.  We find
that models with relatively large voids on the last scattering surface
predict too much CMB power in an Einstein--de Sitter background
cosmology but could be consistent with the current CMB observations in
a $\Lambda$CDM universe.}

\section{Introduction}
Analyses of surveys such as the 2-degree Field Galaxy Redshift Survey
and the Sloan Digital Sky Survey indicate large volumes of relatively
empty space, or voids, in the distribution of
galaxies~\cite{hoy02,hoy04}. In the standard hierarchical model of
structure formation, gravitational clustering is responsible for
emptying these voids of mass and galaxies~\cite{pee89}. However,
standard cold dark matter (CDM) model simulations predict significant
clumps of matter within voids that are capable of developing into
observable bound objects and are not seen in the
surveys~\cite{dek86}$^-$\cite{sha04}.  Peebles gives an in-depth
discussion of the contradictions of this prediction with
observation~\cite{pee01}.  He argues that the inability of the CDM
models to produce the observed voids constitutes a true crisis for
these models. Additionally, deep field observations from the Cosmic
Background Imager~\cite{mas03} (CBI) show excess power on small
angular scales, $\ell > 2000$, in the cosmic microwave background
(CMB).

It may be possible to explain these 2 observations by postulating the
presence of a void network originating from primordial bubbles of true
vacuum that nucleated during inflation~\cite{la91,lid91}. In this
scenario, the first bubbles to nucleate are stretched by the remaining
inflation to cosmological scales. The largest voids may have had
insufficient time to thermalise before decoupling and may persist to
the present day.  Such primordial voids are predicted to produce a
measurable contribution to the CMB~\cite{gri03}.

In this paper, we discuss the results of Griffiths {\it et al.}
(2003)~\cite{gri03} in which we develop a general method to
approximate the void contribution to the CMB that allows the creation
of maps and enables us to consider an arbitrary distribution of void
sizes. We show that if the voids that we see in galaxy surveys today
existed at the epoch of decoupling, they could contribute significant
additional power to the CMB angular power spectrum between $2000 <
\ell < 3000$. Further to this work, we describe how we improve our
predictions. We include the effects of the cosmological constant as
well as oscillations in the matter-radiation fluid that may be
generated by primordial voids on scales up to the sound horizon.  For
full details of our improved methodology refer to our recently
submitted manuscript on the subject~\cite{ord05}.

\section{First Approximation}
\subsection{Void parameters}
We model the voids seen today as spherical underdensities of $\delta
\rho/\rho = -1$.  Each void is bounded by a thin wall containing the
matter that is swept up during the void expansion.  This forms a
compensated void. This means that the overall cosmology is that of the
background universe, since a compensated void does not distort
space-time outside of itself. This is an extremely important property,
as it allows us to place many voids into a universe without the worry
that they might influence each other.  As a first approximation we
take the background universe to be an Einstein--de Sitter (EdS)
cosmology. Conservation of momentum~\cite{mae83} or
energy~\cite{ber85} can be used to show that compensated voids in an
EdS background cosmology will increase in radius $r_v$ between the
onset of the gravitational collapse of matter at equality and the
present day such that
\begin{equation}
\label{e:vexpand}
r_v(\eta) \propto \eta^{\beta} \,,
\end{equation}
with $\beta \approx 0.39$ and where $\eta$ is conformal time.

Motivated by the extended inflationary scenario
\cite{las89,kol91}, we assume a power-law distribution of bubble
sizes greater than a given radius $r$ of the form,
\begin{equation}
N_B(>r) \propto r^{-\alpha} . \label{eqvoidsize}
\end{equation}
Typically, extended inflation is implemented within the framework of a
Jordan-Brans-Dicke theory \cite{bra61}. In this case, the exponent
$\alpha$ is directly related to the gravitational coupling $\omega$ of
the scalar field that drives inflation,
\begin{equation}
\alpha = 3 + \frac{4}{\omega + 1/2} .
\end{equation}
Values of $\omega>3500$ are required by solar system
experiments~\cite{wil01}. We take the limit of large $\omega$, leading
to a spectrum of void sizes with $\alpha = 3$.

It has been shown that a distribution of void sizes that predicts the
existence of arbitrarily large inflationary voids will cause
significant effects on the CMB that contradict
observations~\cite{lid91}.  We assume that the mechanism creating the
voids imposes an upper cut-off on the size distribution. A possible
mechanism for this cut-off could be that the tunneling probability of
inflationary bubbles is modulated through the coupling to another
field.

As well as avoiding the well known problems associated with
arbitrarily large voids existing at the epoch of decoupling, this
assumption allows us to match the observed upper limit on void sizes
from the galaxy redshift surveys. We choose the average value that is
found, $r_{\rm max} = 25$ Mpc$/h$
\cite{hoy02,hoy04}. The minimal present void size is also
chosen to agree with redshift surveys, $r_{\rm min}=10$ Mpc$/h$.

We normalise the distribution by requiring that the total number of
voids satisfies the observed fraction of the universe filled with
voids today, $F_v$.  Redshift surveys indicate that approximately 40\%
of the fractional volume of the universe is in the form of voids of
underdensity $\delta \rho/\rho < -0.9$ \cite{hoy02,hoy04}, ie. $F_v
\approx 0.4$. The positions of the voids are then assigned randomly,
making sure that they do not overlap. In order to speed up this
process, we consider only a $10^\circ$ cone. This limits our analysis
to $\ell > 100$, which is satisfactory for our purpose since the main
contribution from voids is on much smaller scales.

\subsection{Stepping through the void network}\label{sec:vstep}
Refer to Griffiths {\em et al.} (2003)~\cite{gri03} for a more detailed
description of our methodology.  We ray trace photon paths from us to
the last scattering surface (LSS) for the 10$^\circ$ cone in steps of
1'.  Each void in the present day distribution that is intersected by
the photon path is evolved back in time according to
equation~(\ref{e:vexpand}) to determine whether the photon encounters
the void.

\begin{figure}[t]
\begin{center}
\psfig{figure=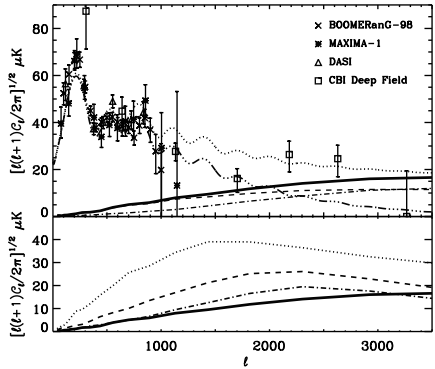,height=2in}
\caption{Top: The CMB anisotropies produced by the fiducial EdS void
model (solid line) compared to the primary $\Lambda$CDM CMB
anisotropies (dashed-triple-dotted). Also plotted are the sum of
primary and void contributions (dotted) as well as the fluctuations
induced purely by voids on the last scattering surface (dashed) and by
those between last scattering and today (dashed-dotted).  We show the
``standard'' cosmological concordance model: of course a combined
analysis of primary and void--induced fluctuations would select a
different cosmology for the primary contribution.
Bottom: Example models depicting a range of void contributions to the 
CMB fluctuations. The models plotted are 
$\alpha = 3$, $r_{\rm max}=25$ Mpc$/h$ and
$F_v=0.4$ (solid line), $\alpha = 3$, $r_{\rm max}=40$ Mpc$/h$ and
$F_v=0.4$ (dotted), $\alpha = 3$, $r_{\rm max}=40$ Mpc$/h$ and
$F_v=0.2$ (dashed) and $\alpha = 6$, $r_{\rm max}=40$ Mpc$/h$ and
$F_v=0.4$ (dashed-dotted).}\label{figcl}
\end{center}
\end{figure}

If a photon intersects a void between us and the LSS, we compute the
Rees--Sciama (RS) effect~\cite{rs68} due to the deviation in the
redshift of the photon as it passes through the expanding void and the
lensing effect due to the deviation in its path. If a photon
intersects a void on the LSS, we calculate the Sachs--Wolfe (SW)
effect~\cite{sw67} due to the photon originating from within the
underdensity. We take into account the finite thickness of the LSS,
which suppresses the SW effect for small voids, by averaging the
contribution from a number of photons originating from a LSS of mean
redshift 1100 and standard deviation in redshift 80.  We also
calculate the partial RS effect that arises due to the expansion of
the void on the LSS as the photon leaves it.

Once the photon has reached the last scattering surface, we know the
variation of its temperature as well as its position on the LSS and
can create a temperature map.  We then use a flat sky approximation
\cite{whi99,das02} to obtain the $C_\ell$ spectrum of the anisotropies
(see Fig. \ref{figcl}). We point out that primordial void parameters
are still poorly constrained by both observation and theory.  The
bottom panel of Fig.~\ref{figcl} shows a few further example models.

For a power-law size distribution (as motivated by the inflationary
scenario), large voids become rarer as $\alpha$ is increased.
Therefore, since void analyses of redshift surveys only sample a
fraction of the volume of the universe, there may exist voids of
larger $r_{\rm max}$ than currently observed.  Models with high
$r_{\rm max}$ voids in an EdS background cosmology tend to predict too
much power on scales $\ell \approx 1000$. However, if we take
inflationary models with $\alpha > 6$ then the peak moves to larger
$\ell$ and the total power drops. The filling fraction mainly adjusts
the overall power.


\section{Improved Prediction}
\subsection{Void Evolution in a $\Lambda$CDM background cosmology}
The addition of a cosmological constant ($\Lambda$) to the cosmology
is expected to slow down the conformal evolution of the voids at late
times.  We test this hypothesis by simulating the comoving evolution
of a single void of radius $25$ Mpc$/h$ starting at $z=1000$ in a
$(100$ Mpc$/h)^3$ box containing $64^3$ particles.  We find
that the void size evolution in a $\Lambda$CDM background deviates
from that of the EdS scaling solution at late times as expected.
However, the final radius is underestimated by less than 2\%.  The EdS
void scaling relationship given by equation~(\ref{e:vexpand}) with
$\beta = 0.39$ is therefore a good approximation for a void in a
$\Lambda$CDM background.

\begin{figure}
\centering
\psfig{figure=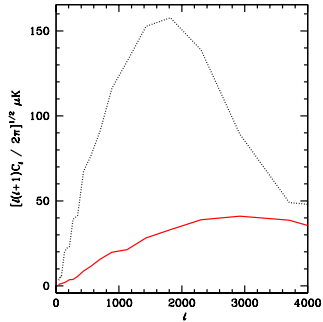,height=1.8in}
\caption[equally sized voids]{\label{fig:lamveds} The predicted CMB
  power from a uniform distribution of equally sized voids in EdS
  (dotted) and $\Lambda$CDM (solid) background universes. The voids have a
  size of 40 Mpc/$h$ and a filling fraction of $40$ \%.  The void
  contribution is computed by ray tracing a void network.  The effect
  of $\Lambda$ is to suppress the effect of the voids and move the
  peak in the power spectrum to smaller angular scales since they will
  appear smaller on the last scattering surface.}
\end{figure}

Since a $\Lambda$CDM universe evolves for substantially longer, we
would expect the voids that we see today to appear smaller on the last
scattering surface for this cosmology than in our first approximation
EdS background. We would therefore predict that voids in a
$\Lambda$CDM universe will have a more suppressed effect on the CMB
than we have first approximated. This can be modelled by taking the
horizon size of our cone of voids to be that of a $\Lambda$CDM
cosmology. Fig.~\ref{fig:lamveds} compares the predicted power from
a uniform distribution of equally sized voids in EdS and $\Lambda$CDM
background cosmologies.  The overall contribution to the CMB from the
voids is suppressed as expected and the peak is moved to smaller
angular scales since the voids now appear smaller on the last
scattering surface.

\subsection{Modelling acoustic oscillations}
So far we have studied voids which do not interact with their
surroundings. Specifically, they do not perturb the matter-radiation
fluid themselves. While this is a reasonable approximation for the RS
effect after recombination, it has been pointed out \cite{bac00} that
this is not true before last scattering. A void could therefore set up
oscillations in the matter-radiation fluid on scales up to the sound
horizon which might be clearly visible in the power spectrum.

\begin{figure}[t]
\begin{center}
\psfig{figure=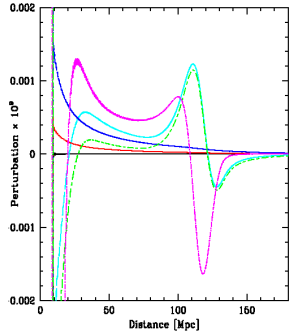,height=1.8in}
\caption{A ``zoom'' onto the sound horizon showing the behaviour of
the perturbations in real space at last scattering (LS) from a single
void. The void, which has a radius of 40 Mpc today, has a size of 9
Mpc at LS. The black solid curve is the input perturbation
$\Phi_S(x,\eta_{LS})/2$, while the red dotted curve is the total
metric perturbation $\Phi$/2. The blue small dashed curve shows
$-2\Psi=2\Phi$. The cyan large dashed curve is the temperature
perturbation from the photons, $D_g/4$.  The baryon velocity is shown
as the magenta dot-dashed curve.  The green small dashed-large dashed
curve is $D_g/4-2 \Phi$, which is very similar to $\Phi_S/2$.
}\label{fig:acoustic}
\end{center}
\end{figure}

The standard way to include the full behaviour of matter and radiation
perturbations is to solve the Boltzmann equation numerically. But it
is not possible to model a void self-consistently inside the Boltzmann
solver. A similar problem was encountered a few years ago while
investigating topological defects~\cite{dur02}. As opposed to defects,
which are entirely external sources, the voids represent a
perturbation of the dark matter itself. We have chosen to model this
by making the approximation that the dark matter is completely
decoupled from the rest of the universe. Therefore, the cold dark
matter acts only as an external seed where the perturbations are
concerned (of course it is taken into account for the background
quantities). We insert our source into a modified version of CMBEasy
\cite{dor03} and then write out a snapshot of the fluids at the time
of last scattering and Fourier transform them back.

\begin{figure}
\centering
\psfig{figure=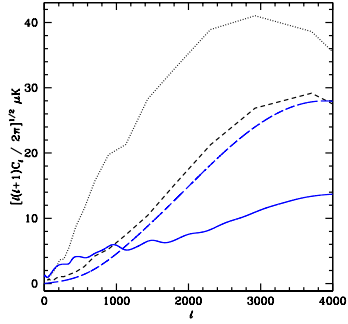,height=1.8in}
\caption[equally sized voids]{\label{fig:compare} We compare the
  $C_\ell$ from ray tracing and the Boltzmann approach. The dotted
  curve shows the total $\delta T$ for a $\Lambda$CDM model with voids
  of a size of $40$ Mpc$/h$ and a filling fraction of $40$ \%, computed
  by ray tracing a void network. The solid curve shows the same but
  computed with a Boltzmann code. The small and large dashed lines show
  the RS effect due to $\Phi_v$ alone for both
  methods and with the same approximations, these two curves should coincide.}
\end{figure}

Fig.~\ref{fig:acoustic} shows a close up of the perturbations for our
model of void formation around the sound horizon. We see that the
amplitude of the sound waves is only about 1\% of the amplitude of the
temperature perturbations inside the void. For smaller voids, the
relative amplitude will be even less.  This seems quite unimportant,
however, the void itself covers a surface of only about $9^2 \pi \, ({\rm
Mpc}/h)^2$ on the LSS, the sound horizon in our flat matter dominated
universe of the example case is at $111 {\rm \, Mpc}/h$ and the sound
waves extend out to about $100 {\rm \,  Mpc}/h$, ten times further than
the size of the void. Therefore, the total power in the fluctuations
is approximately $100$ times larger than expected from their amplitude
and is comparable to the the power from the void predicted by our ray
tracing method. Indeed, we see oscillations appear on large angular
scales in the angular power spectrum (see Fig.~\ref{fig:compare}). For
further details of our method, please refer to Ord {\em et al.}
(2005)~\cite{ord05}.

\section{Summary}
Fig.~\ref{fig:compare} compares the void $C_\ell$ from ray tracing and
the Boltzmann approach for a uniform distribution of $40$ Mpc/$h$ sized
voids.  The overall effect from the first approximation ray tracing
method appears to be suppressed using the improved Boltzmann
solution. Evidence of acoustic oscillations generated in the
photon-baryon fluid can also be seen on large angular scales. The RS
effect due to $\Phi_v$ alone, using the same approximations in both
calculations, is in agreement for both methods as expected.

The suppression that is evident using the Boltzmann solution over the
ray tracing approach is due to a number of factors.  In the ray tracing
method we assume that the density contrast $\delta=-1$ and we take
$(a'/a)^2$ to be $4/\eta^2$.  Close to radiation domination (ie. at
early times, just after last scattering, when we get the biggest
effect) both these approximations act to increase the ray tracing
result with respect to the more accurate Boltzmann prediction,
together by more than a factor of 2. The ray tracing method also
neglects the Doppler contribution from the baryon velocity, which
suppresses the result further.  Finally, the finite thickness of the
LSS and Silk damping were not fully taken into account in our first
approximation ray tracing method.

\begin{figure}[t]
\centering
\psfig{figure=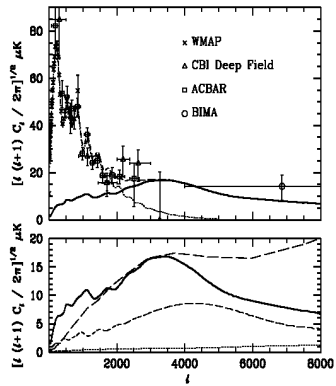,height=2in}
\caption[equally sized voids]{\label{fig:final}Top: The CMB
anisotropies produced by a $\Lambda$CDM void model (solid line:
$\alpha = 3$, $r_{\rm max}=55$ Mpc$/h$ and $F_v=0.4$) compared
to the primary $\Lambda$CDM anisotropies (dotted). Also plotted are
the sum of primary and void contributions (dashed). We show the
``standard'' cosmological concordance model. Again a combined analysis
of primary and void--induced fluctuations would select a different
cosmology for the primary contribution.  Bottom: The large dashed
curve shows the fiducial EdS model of Fig.~\ref{figcl} ($\alpha =
3$, $r_{\rm max}=25$ Mpc$/h$ and $F_v=0.4$) using our ray
tracing method. The same model is shown for our Boltzmann solution
with EdS (small dashed) and $\Lambda$CDM (dotted) background
universes.  The solid bold line is the same $\Lambda$CDM void model
for $r_{\rm max}=55$ Mpc$/h$ maximally sized voids as in the
panel above.  }
\end{figure}

As discussed in subsection~\ref{sec:vstep}, there may exist voids of
larger $r_{\rm max}$ than currently observed.  Models with $\alpha=3$
and high $r_{\rm max}$ voids in an EdS background cosmology tend to
predict too much power on scales $\ell \approx 1000$.  Our results
show that this is less of a problem for a $\Lambda$CDM
universe. Furthermore, since the Boltzmann solution predicts the CMB
power from voids to be lower than implied by the ray tracing
estimation, relatively large voids on the last scattering surface may
be consistent with the current CMB data, even for void distributions
with low values of $\alpha$ (see Fig.~\ref{fig:final}). 

Experiments such as CBI are able to directly probe small angular
scales and constrain void parameters. We will present a Markov Chain
Monte Carlo constraints analysis of a wide range of void models in a
future paper.  We will further constrain models that are compatible
with CMB observations using cluster evolution \cite{mat04} and also
investigate the non-Gaussian signal of void models that are compatible
with the observations.

\section*{Acknowledgments}
We thank Michael Doran for a prerelease version of his CMBEasy code and
his support in modifying it. It is a pleasure to thank Ruth Durrer and
Andrew Liddle for helpful discussions and comments.  LMO acknowledges
support from ARC.  MK acknowledges support by PPARC and the Swiss
Science Foundation.

\end{document}